\documentclass[1p,12p]{elsarticle}

\usepackage{titlesec}
\titleformat*{\section}{\LARGE \bfseries}
\titleformat*{\subsection}{\Large \bfseries}
\titleformat*{\subsubsection}{\large \bfseries}
\titleformat{\paragraph}
{\normalfont\normalsize\bfseries}{\theparagraph}{1em}{}
\titlespacing*{\paragraph}{0pt}{3.25ex plus 1ex minus .2ex}{1.5ex plus .2ex}

\usepackage{multirow}
\usepackage{titlesec}
\usepackage{amssymb}
\usepackage{moreverb}
\usepackage{empheq}
\usepackage{amssymb,bm,amsbsy,graphicx,color,amsmath,amsfonts}
\usepackage{mathrsfs}
\usepackage{latexsym}
\usepackage{pseudocode}
\usepackage{epsfig,epsf,rotating}
\usepackage{subfig}
\usepackage{epstopdf}
\usepackage{framed}
\usepackage[svgnames]{xcolor}
\usepackage{alltt}
\usepackage{lipsum}
\usepackage{mathrsfs}
\usepackage{tikz}
\usepackage[utf8]{inputenc}
\usepackage[T1]{fontenc}
\usepackage{ae,aecompl}
\usepackage[font=normalsize]{caption}
\usepackage{listings}
\usepackage[labelfont=bf]{caption}
\usepackage{nameref}
\usepackage{graphicx}
\usepackage{algpseudocode}
\usepackage{algorithm}
\usepackage{makecell}
\usepackage{pgf-pie}
\usepackage{booktabs}
\usepackage{multirow}
\usepackage{siunitx}

\DeclareMathAlphabet{\mathscrbf}{OMS}{mdugm}{b}{n}
\DeclareCaptionType[fileext=ext]{Video}
\biboptions{sort&compress}
\newcommand{\vect}[1]{\boldsymbol{#1}}

\usepackage[colorlinks,bookmarksopen,bookmarksnumbered,citecolor=red,urlcolor=red]{hyperref}

\def\code#1{{\texttt{#1}}}
\definecolor{codegreen}{rgb}{0,0.6,0}
\definecolor{codegray}{rgb}{0.5,0.5,0.5}
\definecolor{codepurple}{rgb}{0.58,0,0.82}
\definecolor{backcolour}{rgb}{0.95,0.95,0.92}

\makeatletter
\def\ps@pprintTitle{%
	\let\@oddhead\@empty
	\let\@evenhead\@empty
	\def\@oddfoot{\reset@font\hfil\thepage\hfil}
	\let\@evenfoot\@oddfoot
}
\makeatother

\usepackage{lipsum} 
\usepackage{fancyhdr}
\begin{document}

\begin{frontmatter}

\title{\LARGE \textbf{Machine learning and reduced order modelling for the simulation of braided stent deployment.}}

\author[1,2,3]{Beatrice Bisighini} \author[4,5,1]{Miquel Aguirre} \author[3]{Marco Evangelos Biancolini} \author[3]{Federica Trovalusci} \author[2]{David Perrin} \author[1]{Stéphane Avril \corref{cor1}} \author[1]{Baptiste Pierrat}

\cortext[cor1]{Corresponding author: Stéphane Avril (\href{mailto:avril@emse.fr}{avril@emse.fr}).}

\address[1]{Mines Saint-Etienne, Univ Jean Monnet, INSERM, U 1059 Sainbiose, Centre CIS, F-42023 Saint-Etienne, France}
\address[2]{Predisurge, 10 rue Marius Patinaud, Grande Usine Creative 2, 42000 Saint-Etienne, France}
\address[3]{University Tor Vergata, Department of Enterprise Engineering, Via del Politecnico 1, 00133 Rome, Italy}
\address[4]{Laboratori de Càlcul Numèric, Universitat Politècnica de Catalunya, Jordi Girona 1, E-08034, Barcelona, Spain}
\address[5]{International Centre for Numerical Methods in Engineering (CIMNE), Gran Capità, 08034, Barcelona, Spain}

\large{\begin{abstract}
Endoluminal reconstruction using flow diverters represents a novel paradigm for the minimally invasive treatment of intracranial aneurysms. The configuration assumed by these very dense braided stents once deployed within the parent vessel is not easily predictable and medical volumetric images alone may be insufficient to plan the treatment satisfactorily. Therefore, here we propose a fast and accurate machine learning and reduced order modelling framework, based on finite element simulations, to assist practitioners in the planning and interventional stages. It consists of a first classification step to determine a priori whether a simulation will be successful (good conformity between stent and vessel) or not from a clinical perspective, followed by a regression step that provides an approximated solution of the deployed stent configuration. The latter is achieved using a non-intrusive reduced order modelling scheme that combines the proper orthogonal decomposition algorithm and Gaussian process regression. The workflow was validated on an idealised intracranial artery with a saccular aneurysm and the effect of six geometrical and surgical parameters on the outcome of stent deployment was studied. The two-step workflow allows the classification of deployment conditions with up to 95$\%$ accuracy and real-time prediction of the stent deployed configuration with an average prediction error never greater than the spatial resolution of 3D rotational angiography (0.15 mm). These results are promising as they demonstrate the ability of these techniques to achieve simulations within a few milliseconds while retaining the mechanical realism and predictability of the stent deployed configuration.
\end{abstract}}

{\large }

\end{frontmatter}

\section{Introduction}

Intracranial aneurysms (IAs) are local dilations of the arteries in the brain caused by a degenerative weakening of the arterial wall. Saccular, or blister-like, are the most common IAs. Their prevalence among the general population is estimated to be around 2-3$\%$ \citep{Rinkel1998}. With an incidence of 10/100000 person-years, IAs rupture leads to subarachnoid haemorrhage, a life-threatening type of stroke with high morbidity and mortality \citep{Vlak2011, D.O.Wiebers2003}. Therefore, when IAs with a diameter larger than 5/7 mm are detected, they are often recommended for early treatment to exclude the aneurysm sac from the cerebral circulation \citep{Rahman2010}. Nowadays, endovascular options have become the preferred intervention thanks to the lower rate of complications with respect to invasive techniques (e.g., clipping) \citep{Liu2015}. 

Endoluminal reconstruction through flow diverters represents a novel paradigm for the mini-invasive treatment of IAs, as an alternative to endosaccular occlusion through coiling \citep{Pierot2011, Durso2011}. Flow diverters are self-expanding devices consisting of a low-porosity braided stent. Thanks to this structure, they are highly flexible and resistant to kinking. Accordingly, they are well suited for tortuous vessels and wide-neck IAs. These stents come in different sizes, in terms of radius and length. Nowadays, surgeons choose the size based only on clinical experience and measurements taken on medical volumetric images (e.g., 3D rotational angiography or computed tomography angiography), acquired shortly before surgery. 

However, the configuration assumed by these devices once deployed within the parent vessel is not easily predictable and routine 3D medical images alone may be insufficient to plan the treatment satisfactorily. The related difficulties can lengthen the interventional times. Moreover, higher doses of angiographic contrast agents and anaesthetic drugs are needed. All these factors increase the risk of postoperative complications for the patient. Therefore, there is a compelling need to develop computational models capable of simulating, in real time, the deployment of flow diverters within patient-specific vessels to assist practitioners in the planning and interventional stages \citep{Karmonik2005}.

The mechanical behaviour of braided stents is typically simulated using a finite element (FE) model where beam elements are used to discretise the wires \citep{Auricchio2011, Zaccaria2020, McKenna2021, Shiozaki2020}; however, only a few studies modelled numerically flow diverters \citep{Ma2012, Fu2017, Ma2013}. Due to the large amount of degrees of freedom (DOFs) and the necessity to solve the contact between the device and the vessel wall, the computational time required by these traditional techniques is very high. To overcome this limitation and make computational models suitable for clinical use in real time, fast virtual stenting (FVS) methods have been reported in the literature \citep{Larrabide2012, Zhong2016, Spranger2015a}. They predict the stent deployed configuration by simplifying its mechanical behaviour and/or the contact against the vessel wall (e.g., by using mass-spring models or by using active contour models).

Reduced order modelling is also gaining interest in computer-aided surgery thanks to its capability of reducing the computational complexity and cost of numerical problems while preserving their inner physics \citep{ Niroomandi2017, Mena2015, Niroomandi2012, Santo2020}. One of the most powerful and widespread techniques to build reduced-order models (ROMs) is the reduced-basis (RB) method. The RB method is adapted to non-linear problems which need to be solved a large number of times for different parameter values \citep{Hesthaven2015, Quarteroni2015}. In biomechanics, the parametrization can concern boundary and initial conditions \citep{Girfoglio2022, Chang2017, Bridio2022}, loads \citep{Biancolini2018} and the geometrical domain under investigation \citep{Kardampiki2022, Ballarin2015c, Biancolini2020g}. The latter is sometimes handled with a statistical shape model, which enables the comparison of different anatomies in terms of the same characteristics \citep{Buoso2021, Lauzeral2019a, Cosentino2020}. 

The real-time simulation of such parametrized problems is achieved by exploiting the intrinsic similarities between their solutions. The most important features of the original, full-order model (FOM), i.e. the RBs, can be extracted through proper orthogonal decomposition (POD) from a set of high-fidelity (HF) solutions. The construction of such a dataset and the extraction of the RBs is referred to as \emph{offline} stage. Thereafter, approximated solutions for unseen parameter values are determined as linear combination of the RBs (\emph{online} stage). The powerful advantage of these methods is that, if the \emph{online} stage is completely decoupled from the \emph{offline} one, the computations performed in the \emph{online} stage are independent of the dimension of the FOM. RBs-based methods differ in the implementation of the \emph{online} stage: intrusive methods rely on a projection onto the RBs space to generate the ROM \citep{Ballarin2016b, Ballarin2015b}; non-intrusive methods employ a regression model trained to learn the mapping from parameters to the solution expressed in the RBs space on the HF dataset \citep{Guo2018, Guo2019, Hesthaven2018a, Fresca2021b}. Non-intrusive methods outperform intrusive methods in terms of efficiency, as they do not require solving a system of non-linear equations in the online phase, but only evaluating the regression model. However, they require a very large training dataset to ensure accurate solutions. Recently, physics-informed neural networks (PINNs) emerged as a promising alternative \citep{Chen2021, Buoso2021, Liu2020}.

\begin{figure}[ht!]
	\centering
	\includegraphics[width=1\textwidth]{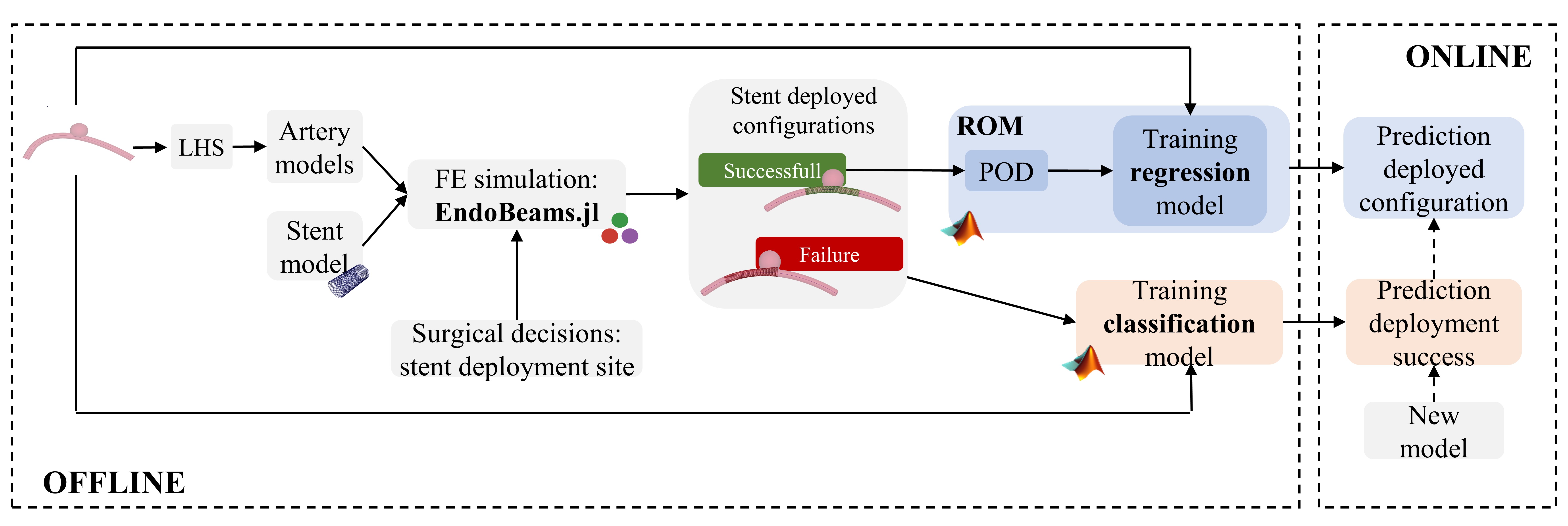}
	\caption{Virtual workflow for the prediction of the outcome of stent deployment simulations.}
	\label{fig:schema}
\end{figure}

In this study, we present the first implementation of a machine learning (ML)-based ROM scheme for the prediction of the stent deployed configuration. To assess its feasibility, we created a parametric synthetic geometry that allows control of the vessel radius, curvature and aneurysm size. Surgical decisions on the deployment site are also considered when creating the HF dataset. If given the deployment conditions the stent does not land inside the aneurysm and is well positioned against the vessel wall, the deployment is considered successful from a clinical perspective. Since there is no clinical need to predict the stent configuration after an unsuccessful deployment, the virtual framework here proposed consists of two steps (Figure \ref{fig:schema}): a first \emph{classification} step that allows a priori determination of whether a simulation will be successful or not, followed by a \emph{regression} step that provides an approximated solution of the deployed stent configuration. Moreover, in continuation with our previous work \citep{Bisighini2022}, we propose a fast strategy to perform FE simulations of braided stent deployment to reduce the computational time required to build the HF dataset for training.

The remainder of the paper is organized as follows. Section \ref{sec:hf} summarizes the methods used to generate the HF dataset with particular emphasis on the description of the flow diverter model (\ref{sec:stent}), the synthetic aneurysm model (\ref{sec:artery}) and the scheme followed to perform FE simulations of stent deployment (\ref{sec:stentsim}). Section \ref{sec:class} describes the classification model, Section \ref{sec:rom} the ML-based ROM. In Section \ref{sec:results} and \ref{sec:disc}, the results of testing these models against unseen scenarios are shown and discussed. Finally, Section \ref{sec:conclusions} presents some concluding remarks.

\section{Methods}

\subsection{High-fidelity simulations}\label{sec:hf}
\subsubsection{Braided stent modelling}\label{sec:stent}
\begin{figure}[ht!]
	\centering
	\includegraphics[width=\textwidth]{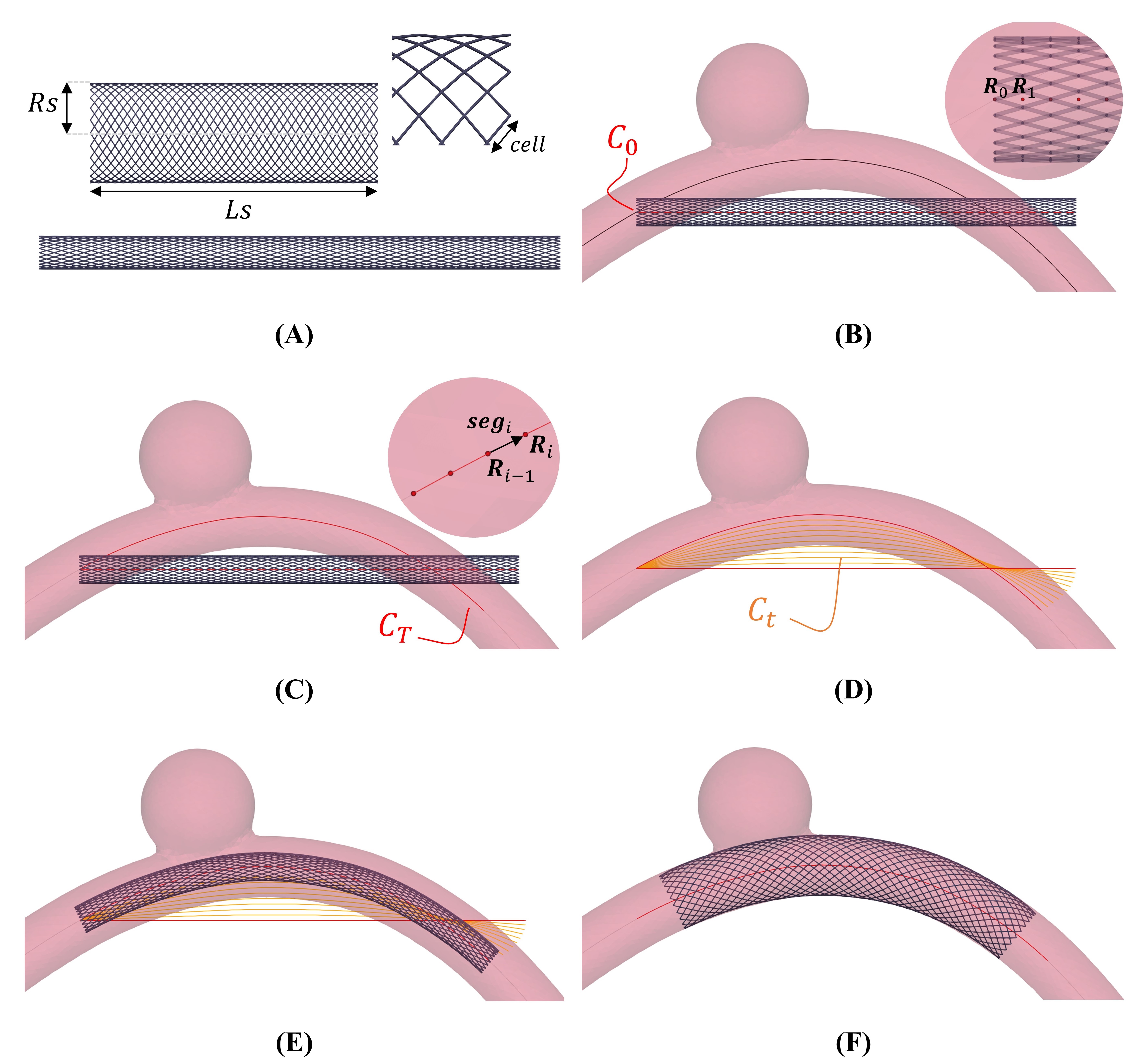}
	\caption{Stent deployment strategy. The wire thickness is magnified (2$\times$) to better visualise the stent. \textbf{(A)} Free and crimped stent. \textbf{(B)} Crimped stent and initially straight centerline ($C_0$). \textbf{(C)} Final centerline of the stent ($C_T$). \textbf{(D)} Intermediate centerlines ($C_t$). \textbf{(E)} Positioned stent. \textbf{(F)} Deployed stent.}
	\label{fig:stent_sim}
\end{figure}

\noindent The braided stent is modelled as a tubular net of thin wires with circular cross-section (Figure \ref{fig:stent_sim}\textbf{A}). For a stent with radius $R_s$, length $L_s$, composed by $N_w$ wires with radius $R_w$ and presenting $N_{cells}$ repetitive units, the nodal positions are defined by the following set of equations: 
\begin{align}
	x_{n,i} &= (R_s + R_w) \cdot \cos(orient \cdot i \cdot d\theta + \Theta_n ), \\
	y_{n,i} &= (R_s + R_w) \cdot \sin(orient \cdot i \cdot d\theta + \Theta_n ),\\
	z_{n,i} &= n \cdot \tan(\phi),
\end{align}
with $n \in [0, N_w/2]$ and $i \in [0, N_{cells}]$ and where $orient$ is either 1 or -1 respectively for clockwise and counter-clockwise wires, $d\theta = {2\pi}/(N_w/2)$, $\Theta_n = n\cdot{2\pi}/(N_w/2)$ and $\phi = {L_s}/{N_{cells}}$ is the pitch angle. 

In this work, the values assigned to these geometrical parameters refer to a generic flow diverter: $N_w$ = 48, $R_s$ = 2.6 mm, $R_w$ = 0.014 mm, $L$ = 15 mm, $N_{cells}$ = 70. The stent is made of Phynox, a cobalt-chromium alloy, which is modelled as a linear elastic material with Young modulus $E$ = 225 GPa, Poisson coefficient $\nu$ = 0.33 and density $\rho$ = 9.13$\cdot$10$^{3}$ kg/m$^3$.

\subsubsection{Artery and aneurysm modelling}\label{sec:artery}
\begin{figure}[ht!]
	\centering
	\includegraphics[width=\textwidth]{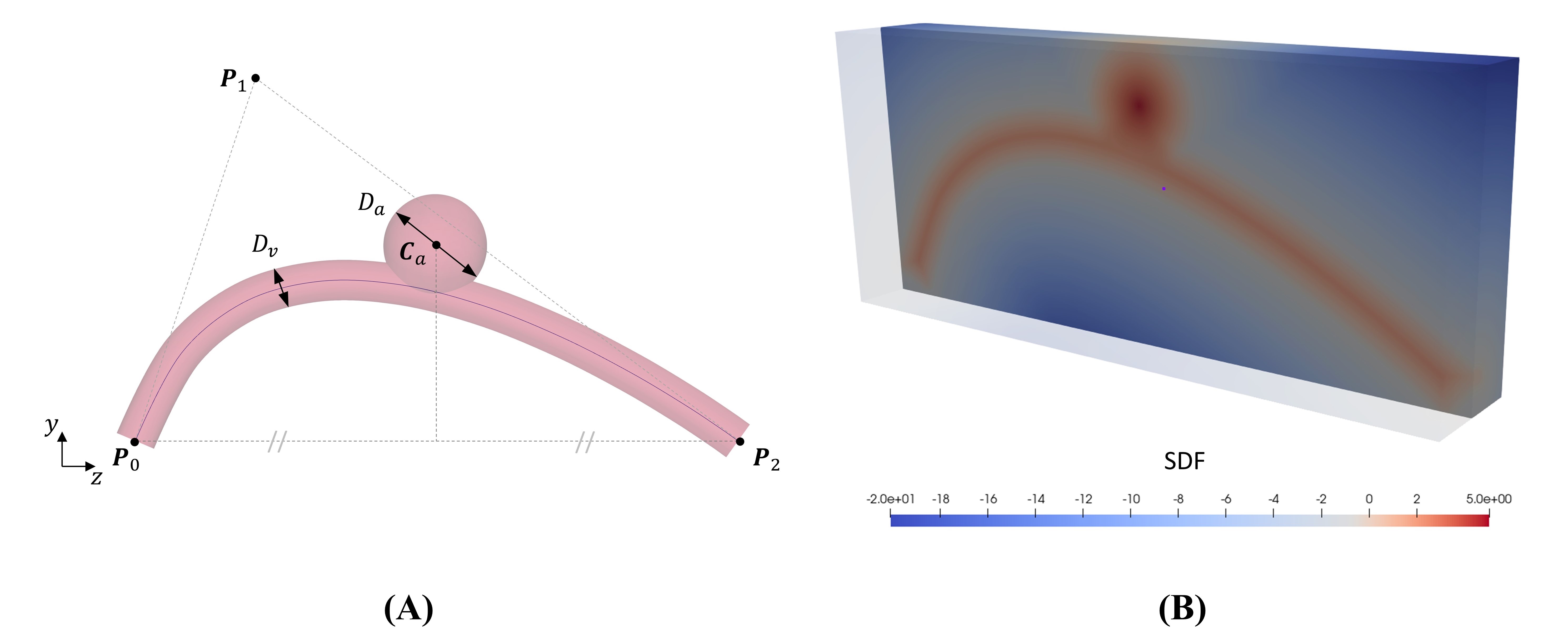}
	\caption{Artery and aneurysm modelling. \textbf{(A)} Parametric idealised model of an intracranial artery presenting a saccular aneurysm: Bézier points $\textbf{P}_0$, $\textbf{P}_1$ $\textbf{P}_2$, artery diameter $D_{v}$, aneurysm diameter $D_{a}$ and centre $\textbf{C}_{a}$. \textbf{(B)} SDF section of the idealised artery an aneurysm model.}\label{fig:vessel_model} 
\end{figure}

\noindent The stent is released within a parametric idealised model of an intracranial artery presenting a saccular aneurysm. The vessel centerline is defined using a planar quadratic Bézier curve:
\begin{equation}
	\textbf{B}(t) = (1-t)^{2}\textbf{P}_0 + 2t(1-t)\textbf{P}_1 +t^2\textbf{P}_2, \quad t\in[0,1],
\end{equation} 
where $\textbf{P}_0$, $\textbf{P}_2$ are fixed and $\textbf{P}_1$ is the control point which will be included in the parametrisation. These points lie in a 2D plane, so the only DOFs of this spline are the $y,z$ coordinates of $\textbf{P}_1$. The curve $\textbf{B}(t)$ starts from $\textbf{P}_0$ in the direction of $\textbf{P}_1$, then bends to reach $\textbf{P}_2$. Bézier curves allow defining smooth, continuous curves that resemble the curvature of intracranial arteries \citep{Zykowski2018, Danu2019}. The Visualization Toolkit (VTK) software is employed to build the model \citep{vtk}. The artery is created using the \code{vtkTubeFilter()} function that generates a tube around a line. Its diameter ($D_{v}$) is considered constant along the centerline. Following, a spherical idealised aneurysm with diameter $D_{a}$ is created using \code{vtkSphereSource()}, the sphere centre $\textbf{C}_{a}$ is positioned in the middle of the vessel centerline and the relative $y$-distance is parametrised. Through a Boolean union, the sphere is added to the artery model (Figure \ref{fig:vessel_model}\textbf{A}). In summary, the vessel geometry is fully parametrised by 5 parameters: $y_{\textbf{P}_1}$, $z_{\textbf{P}_1}$, $D_{v}$, $D_{a}$, $y_{\textbf{C}_a}$.

\subsubsection{FE simulation of braided stent deployment}\label{sec:stentsim}
The simulations are performed using \href{https://gitlab.emse.fr/pierrat/EndoBeams.jl}{EndoBeams.jl}, an in-house and open-source FE modelling framework for the numerical simulation of frictional contact interactions between beams and rigid surfaces \citep{Bisighini2022}. This software has been validated against some public benchmark tests and the commercial software Abaqus (Simulia, Dassault Systems, Providence, RI, USA). Thus, the stent structure is discretised using beam elements and the resulting mesh is composed of 3408 nodes and 3360 beam elements. To avoid rendering beam-to-beam contacts, a penalty-based constraint is imposed at each interconnection between the interlaced wires. The contact algorithm implemented in EndoBeams.jl relies on an implicit representation of the contact surface, a signed distance field (SDF) \citep{Aguirre2020}. The SDF of the artery model is computed using the Julia package SignedDistanceField.jl \citep{SignedDistanceField}, which only requires as input the triangular mesh in the form of a \textit{.stl} file (Figure \ref{fig:vessel_model}\textbf{B}). The vessel wall is assumed rigid, thus the SDF is constant along the simulation.

The strategy followed to perform simulations of stent deployment was inspired by the work proposed in \citep{Spranger2015a, Perrin2015a, Hemmler2018}. This approach is compatible with the use of the SDF to manage the vessel-stent contacts and represents an alternative to the use of a virtual catheter, as typically done before in the literature \citep{Bock2012}. Since our goal is to obtain the stent configuration at the end of the deployment simulation, we focus on quasi-static simulations. The simulations are carried out in three steps: (1) \textit{crimping}; (2) \textit{positioning} and (3) \textit{deployment}.

First, the braided stent is crimped by blocking its circumferential DOFs and imposing a radial displacement equal to $(R_{stent}-R_{crimped})$ to all its nodes, where $R_{crimped}$ is the stent target radius after the crimping (Figure \ref{fig:stent_sim}\textbf{A}).

The braided stent nodes lying on the same $z$-plane can be grouped in $N_r$ rings and a central point can be introduced for each of these rings, $\textbf{R}_i\text{ with } i = 1 \dots N_r$ (Figure \ref{fig:stent_sim}\textbf{B}). These points define the initially straight centerline of the crimped stent ($C_0$). The stent is displaced so that $\textbf{R}_0$ coincides with the chosen deployment site along the vessel centerline. The final centerline of the stent ($C_T$) is computed as projection of $C_0$ along the vessel one (Figure \ref{fig:stent_sim}\textbf{C}) so as to maintain constant the total length of the stent centerline. $C_0$, and so $C_T$, can be subdivided into segments, i.e. vectors connecting subsequent points along the centerline ($\textbf{seg}_i = \|\textbf{R}_i - \textbf{R}_{i-1}\|$). The first point $\textbf{R}_{0}$ is considered aligned with the $z$-axis. A rotation is associated with each of these segments to realign it to the preceding one. The angle ($\theta_i$) and axis ($\textbf{ax}_i$) of rotation for each segment can be computed as follows:
\begin{align}
	\textbf{ax}_i &= \frac{\textbf{seg}_i\times \textbf{seg}_{i-1}}{\|\textbf{seg}_i\|\|\textbf{seg}_{i-1}\|} \\
	\theta_i &= \cos^{-1} \left(\frac{\textbf{seg}_i\cdot \textbf{seg}_{i-1}}{\|\textbf{seg}_i\|\|\textbf{seg}_{i-1}\|}\right)
\end{align}
Thus, a rotation matrix $\mathcal{M}_i$ can be defined for each segment. By cumulatively applying these rotations, we can align $C_T$ with the $z$-axis, obtaining $C_0$:
\begin{align}
	\mathcal{M}_{tot,i} = \prod_{k=1}^{i} \mathcal{M}_i
\end{align}
The advantage of this technique is the possibility of obtaining intermediate positions ($C_t$) between $C_T$ and $C_0$ by dividing $\theta_i$ by the number of desired configurations and applying only this partial rotation to $C_T$ (Figure \ref{fig:stent_sim}\textbf{D}). These intermediate positions are used to drive a physical simulation where kinematics constraints are implemented between all the stent nodes lying on the same ring and their corresponding points $\textbf{R}_i$ on $C_0$; the difference of the coordinates between $C_t$ and $C_0$ is computed and the resulting displacements are applied to each point of $C_0$ in a successive way leading the stent to bend (Figure \ref{fig:stent_sim}\textbf{E}).

Finally, the stent is allowed to freely deform within the vessel and activate the contact against the wall (Figure \ref{fig:stent_sim}\textbf{F}). The simulations are stopped when the kinetic energy falls below a certain threshold ($10^{-12}$ mJ), which represents static mechanical equilibrium. 

\subsubsection{Creation of the high-fidelity dataset}
\noindent The impact of anatomical characteristics and surgical decisions on the final stent configuration is studied. Therefore, we consider a set of geometric features describing both the artery and aneurysm geometry and the stent deployment site along the vessel centerline; we refer to this generic set of parameters as $\textit{simulation}$ parameters, collected in the vector $\vect{\mu}$.

For the creation of the HF dataset, we considered the following simulation parameters: 
\begin{equation}
	\vect{\mu}_B =\left[\begin{array}{cccc} y_{\textbf{P}_1},\quad z_{\textbf{P}_1},\quad D_{v}, \quad D_{a},\quad y_{\textbf{C}_a},\quad \eta \end{array}\right]
\end{equation}
where $y_{\textbf{P}_1}$ and $z_{\textbf{P}_1}$ are the $y$ and $z$ coordinates of the middle Bézier curve point, $y_{\textbf{C}_a}$ the $y$ coordinate of the aneurysm centre point and $\eta$ the stent position along the vessel centerline. Since the impact of stent misplacement is one of the objectives of this study, we consider deployment sites in which one of the ends of the stent falls in the aneurysm neck area.

A Latin hypercube sampling (LHS) method is used to generate $N_s$ different values for $\vect{\mu}_B$; the corresponding artery models are created and a stent deployment simulation is performed within each model as explained in Section \ref{sec:stentsim}. The LHS plan is created using the Julia package LatinHypercubeSampling.jl \citep{lhs}. The simulation parameters are evaluated within a range that resembles that observed in the literature \citep{Krejza2006, Li2013}: for $D_v$, we consider a range of [2,4]mm; for $D_a$, a range of [5,10]mm.

Alternatively, the simulation parameters $\vect{\mu}_B$ are substituted as input value of the ML models with $\vect{\mu}_{cl}$ where, instead of the middle point of the Bézier curve ($\textbf{P}_1$) and the stent deployment site ($\eta$), the $y$ and $z$ coordinates of $N_{cl}$ points ($\textbf{Q}_i$) on the positioned stent centerline ($C_T$) are used:
\begin{equation}
	\vect{\mu}_{cl} =\left[\begin{array}{cccccc} y_{\textbf{Q}_1},\quad z_{\textbf{Q}_1}, \quad \cdots, \quad y_{\textbf{Q}_{N_{cl}}},\quad z_{\textbf{Q}_{N_{cl}}}, \quad D_{v}, \quad D_{a},\quad y_{\textbf{C}_a} \end{array}\right].
\end{equation}
The $\vect{\mu}_{cl}$ vector components are calculated geometrically from the deployment conditions and no FE simulation is required. 

\subsection{Binary classification}\label{sec:class}
\begin{figure}[ht!]
	\centering
	\includegraphics[width=\textwidth]{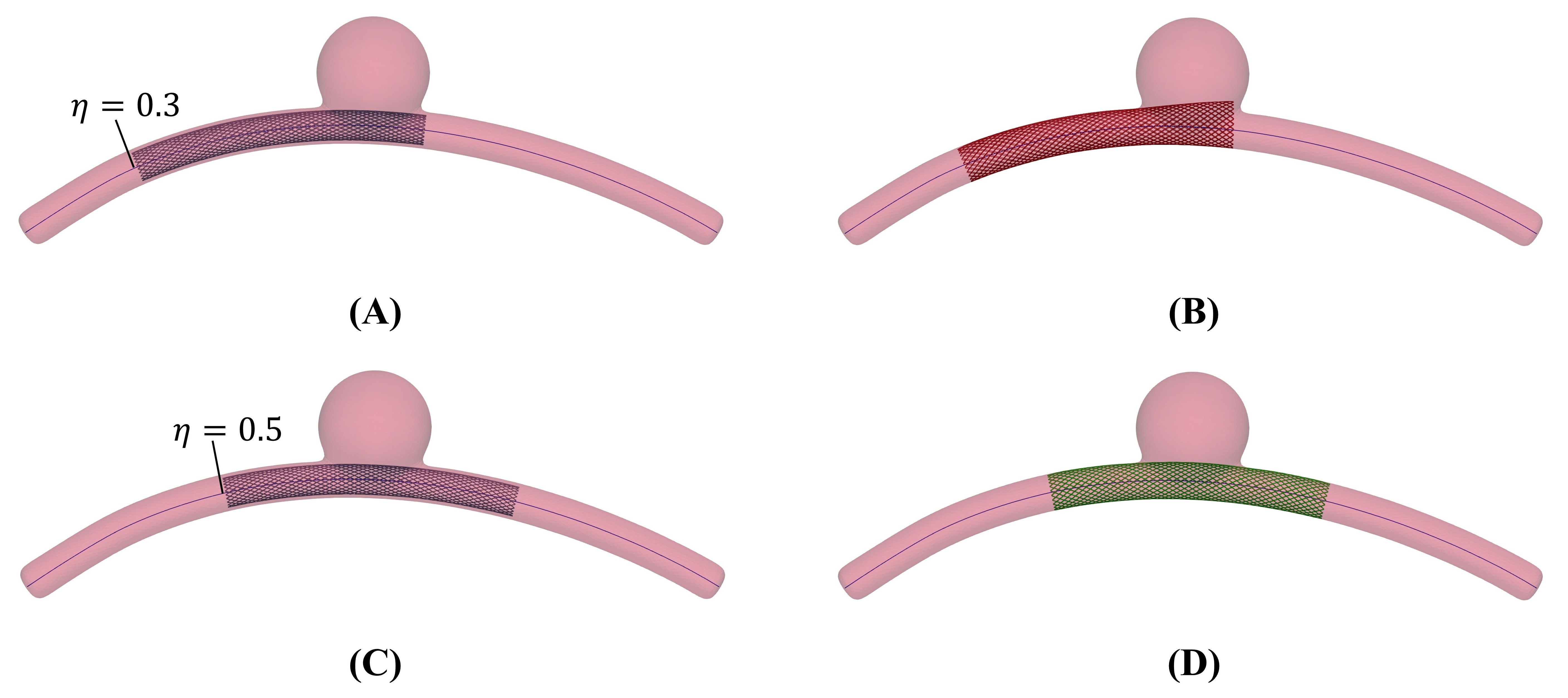}
	\caption{Examples of successful and unsuccessful deployment outcomes depending on the deployment site along the vessel centerline, $\eta$. The wire thickness is magnified (2$\times$) to better visualise the stent. \textbf{(A)} Stent positioned at $\eta = 0.3$. \textbf{(B)} Unsuccessful deployment: landing zone inside the aneurysm space. \textbf{(C)} Stent positioned at $\eta = 0.5$. \textbf{(D)} Successful deployment: good conformity between stent and vessel wall.} \label{fig:class_labels}
\end{figure}

\noindent Within the HF dataset, deployment solutions are considered “successful” from a clinical perspective if the stent extremities are in contact with the artery wall and do not land within the aneurysm sac; otherwise, the simulation outcome is labelled as “failure” (Figure \ref{fig:class_labels}). This classification is done automatically by checking if, at the end of the simulation, one or more nodes from the stent extremities are inside the aneurysm. Being it modelled as a sphere, the SDF of the aneurysm can be computed analytically and, thus, a node $\textbf{x}_{p}$ is located inside the aneurysm if its distance to the surface is positive, i.e.: 
\begin{equation}
	||\textbf{C}_{a}-\textbf{x}_{p}|| - {R}_{a} - {r}_{w} \geq 0.
\end{equation}
Once the results are labelled, a supervised machine-learning algorithm is trained to learn the relationship between the simulation parameters $\vect{\mu}$ and the corresponding solution output: when the model is built, it allows the assignment of new, unseen scenarios to one of the two categories. The classification is done in Matlab (MathWorks, Natick, MA, USA). For this purpose, the performance of the following classifiers is compared \citep{Singh2016}: 
\begin{itemize}
	\item a logistic regression (LR) model is created and fitted using the $\code{fitctree()}$ function; 
	\item a k-Nearest Neighbour (k-NN) model is created and fitted using the $\code{fitcknn()}$ function;
	\item a naive Bayes (NB) model is created and fitted using the $\code{fitcnb()}$ function;
	\item a decision tree (DT) model is created and fitted using the $\code{fitctree()}$ function;
	\item an artificial neural network (ANN) model with three layers of size [10,10,10] and the hyperbolic function $\tanh$ as activation function is created and fitted using the $\code{fitnn()}$ function;
	\item a support vector machine (SVM) model with a polynomial kernel of order 2 is created and fitted using the $\code{fitsvm()}$ function.
\end{itemize}
The architecture and hyperparameters values are chosen for each ML model using the ClassifierLearnerApp.

\subsubsection{Metrics}
Five metrics are considered to evaluate the performance of the trained classifiers. They are computed by counting true positives (TPs), true negatives (TNs), false positives (FPs) and false negatives (FNs), which are collected in the confusion matrix. Their expressions are: 
\begin{itemize}
	\item accuracy = $\frac{\text{TP+TN}}{\text{TP+FN+TN+FP}}$: it tells how good is the classifier, regardless of the label meaning;
	\item sensitivity (or recall or TP rate) = $\frac{\text{TP}}{\text{TP+FN}}$: it tells how good the classifier is at predicting successful deployment conditions;
	\item specificity (or TN rate) = $\frac{\text{TN}}{\text{TN+FP}}$: it tells how good the classifier is at predicting failed deployment conditions;
	\item precision = $\frac{\text{TP}}{\text{TP+FP}}$: it tells how close predicted values are to each other;
	\item F1-score = $2\frac{\text{precision}\cdot\text{recall}}{\text{precision}+\text{recall}}$: it is a more general measure of accuracy that combines precision and recall in a single number.
\end{itemize}
Another useful tool to asses the classification performance is the receiver operating characteristic (ROC), a plot showing the performance of the classifier in terms of TP rate (sensitivity) against FP rate (1-specificity) as a function of the cut-off threshold. The metric related to the ROC curve is the area under the curve (AUC). The closer the ROC curve is to the upper left corner of the graph (and thus the higher the AUC value), the more accurate the classifier is.

\subsection{Reduced order modelling}\label{sec:rom}
\noindent In case of successful deployment, a ML-based reduced order modelling method is employed to compute an approximated solution for the stent deployed configuration within the considered vessel. At first, one might consider training a ML algorithm to learn the relationship between the simulation parameters and the vector of nodal displacements at the end of the simulation. However, given the large number of DOFs of the stent model (in our case, $\sim$10000 DOFs), the size of such an output vector is very large and would result in a very long training time. Reduced order modelling allows for the reduction of the original problem dimension by extracting the most important features, the RBs, from the training dataset through POD \citep{Guo2018, Fresca2022, Han2020b, Bridio2022}. A supervised ML algorithm is then used to establish the relationship between simulation parameters and the solution expressed in the RBs space. Finally, an approximated solution of the stent deployed configuration can be recovered in real time for any combination of simulation parameters.

\subsubsection{Proper orthogonal decomposition}
The theory behind the POD algorithm is here only introduced: if interested, the reader is suggested to refer to \citep{Chatterjee2000}. Once the HF dataset is computed, the snapshots matrix $\textbf{S}$ is built by arranging the HF solutions $\textbf{u}_h(\vect{\mu})$ as columns of a matrix:
\begin{equation}
	\textbf{S} = \left[\begin{array}{cccccccc} \textbf{u}_{h}(\vect{\mu}_1)&|&\textbf{u}_{h}(\vect{\mu}_2)&|& \dots &|& \textbf{u}_{h}(\vect{\mu}_{N_s})\end{array}\right].
\end{equation}
Each of these vectors represents one $\textit{snapshot}$ and contains the nodal displacements at the end of the quasi-static deployment simulation:
\begin{equation}
	\textbf{u}_h(\vect{\mu}) =\left[\begin{array}{cccccccc} {u}_{x,1}, &{u}_{y,1}, &{u}_{z,1}, &\dots, &{u}_{x,N_{n}}, &{u}_{y,N_{n}}, &{u}_{z,N_{n}}\end{array}\right],
\end{equation}
where $N_{n}$ is the number of nodes in the stent mesh, thus the dimension of the HF problem is $N_h = 3 \cdot N_{n}$. 

The POD algorithm relies on performing the singular value decomposition (SVD) of $\textbf{S}$:
\begin{equation}
	\underset{N_h\times N_s}{\textbf{S}} = \underset{N_h\times N_h}{\textbf{U}} \quad
	\underset{N_s\times N_h}{\vect{\Sigma}} \quad \underset{N_s\times N_s}{\textbf{Z}^T},
\end{equation}
where $\textbf{U} = \left[\begin{array}{@{}l@{}} \textbf{u}_{1} | \textbf{u}_{2} | \cdots | \textbf{u}_{N_h} \end{array}\right]$ is the left singular vectors matrix, $\textbf{Z} = \left[\begin{array}{@{}l@{}} \textbf{z}_{1} | \textbf{z}_{2} | \cdots | \textbf{z}_{N_s} \end{array}\right]$ is the right singular vectors matrix and $\vect{\Sigma} = diag(\sigma_1, \sigma_2, \cdots, \sigma_{N_s})$ contains the singular values of $\textbf{S}$, sorted from the largest to the smallest ($\sigma_1 \geq \sigma_2 \geq \cdots \geq  \sigma_{N_s} \geq 0$).

The Schmidt-Eckhart-Young theorem states that the columns of $\textbf{S}$, $\text{Col}(\textbf{S})$, can be well approximated by the first ${L}$ left singular vectors of $\textbf{S}$, i.e. $\text{Col}(\textbf{U})$, if the singular values decay rapidly. Thus, given a tolerance $\varepsilon_{POD}$, $L$ can be found as the minimum integer such that: 
\begin{equation}
	\frac {\sum_{i=1}^{{L}} \sigma_i} {\sum_{i=1}^{N_s} \sigma_i} \ge 1 - {\varepsilon_{POD}}.
\end{equation}
The column vectors $\left[\begin{array}{@{}l@{}} \textbf{u}_{1} | \textbf{u}_{2} | \cdots | \textbf{u}_{{L}} \end{array}\right]$ represent the RBs of the model and are assembled in the matrix $\textbf{V}$.

The HF solution $\textbf{u}_{h}(\boldsymbol\mu)$ can be now projected onto the reduced space defined by $\textbf{V}$:
\begin{equation}
	\textbf{u}_{h}(\boldsymbol\mu) = \textbf{U} \textbf{U}^T \textbf{u}_{h}(\boldsymbol\mu)  \approx  \textbf{V} \textbf{V}^T \textbf{u}_{h}(\boldsymbol\mu) = \textbf{V} \textbf{u}_{{L}}(\boldsymbol\mu) = \textbf{u}_{rb}(\boldsymbol\mu),
\end{equation}
where $\textbf{U} \textbf{U}^T = \textbf{I}$ since $\textbf{U}$ is orthogonal, $\textbf{u}_{{L}}$ are the $L$ projection coefficients associated with the column bases of $\textbf{V}$ and $\textbf{u}_{rb}(\boldsymbol\mu)$ is the solution projected onto the reduced space.

\subsubsection{Gaussian process regression}\label{sec:gpr}
As proposed by Guo in \citep{Guo2018}, Gaussian process regression (GPR) is adopted to approximate the HF solutions for any simulation parameters combination. Only a brief theoretical introduction to single-output GPR is provided here, more details can be found in \citep{Care2018}. The approach employed to extend single-output GPR to multi-output problems is to model each of the outputs independently. This is referred to as independent single-output GPR (IGPR). 

Let $\mathcal{D} = \{(\textbf{x}_i, y_i)\}_{i=1}^N$ be a generic training dataset where $\textbf{x}_i$ represents one input vector and $y_i$ the corresponding output. We assume that ${y}_i$ are noisy observations of an unknown regression function $f(\textbf{x}_i)$: 
\begin{equation}
	y_i = f(\textbf{x}_i) + \varepsilon_i, \quad  \varepsilon_i \sim \mathcal{N}(0, \sigma_i^2).
\end{equation}
GPR consists in calculating the probability distribution over all admissible functions that fit the input data and making predictions when fed with new input data. This is possible by considering $f(\textbf{x}_i)$ distributed as a Gaussian process (GP):
\begin{equation}
	f(\textbf{x}_i) \sim \textnormal{GP}(0, \kappa(\textbf{x}, \textbf{x}')).
\end{equation}
A GP is a collection of random variables, any finite number of which have a joint Gaussian distribution. It is fully described by a mean function (considered 0 for simplicity) and a covariance function $ \kappa(\textbf{x}, \textbf{x}')$, known as \textit{kernel}, which should be chosen carefully according to the problem under study. The most commonly used is the \textit{squared exponential \textnormal{(SE)}} kernel, also known as \textit{radial basis function \textnormal{(RBF)}} kernel:
\begin{equation}
	\kappa(x_i, x_j) = \sigma_\kappa^2\exp\left(-\frac{||x_i-x_j||^2}{2l}\right),
\end{equation}
where the standard deviation $\sigma_\kappa$ and the lengthscale $l$ represent the two kernel hyperparameters.

Following the GP definition, the prior Gaussian distribution for the outputs is: 
\begin{equation}
	\textbf{y}|\textbf{X} \sim \mathcal{N}(0, \textbf{K}_y), \quad \textbf{K}_y = \kappa(\textbf{X}, \textbf{X})+\sigma^2\textbf{I}, 
\end{equation}
where $\textbf{y} = \{y_i, y_2,\cdots,y_N\}$ and $\textbf{X} = [\textbf{x}_1, \textbf{x}_2, \cdots,\textbf{x}_N]$. 

Predictions of the noise-free outputs $\textbf{f}^{*}$ on new input data $\textbf{X}^{*}$ are made by exploiting the joint probability distribution: 
\begin{equation}
	\textbf{f}, \textbf{f}^{*} | \textbf{X}, \textbf{X}^{*}	= \left[\begin{array}{cc} \textbf{f} \\ \textbf{f}^{*} \end{array}\right] \sim \mathcal{N}\left(\left[\begin{array}{cc} \textbf{0} \\ \textbf{0} \end{array}\right], \left[\begin{array}{cc} \textbf{K} & \textbf{K}^{*} \\ \textbf{K}^{*\textnormal{T}} & \textbf{K}^{**} \end{array}\right] \right),
\end{equation}
where $\textbf{K}^{**} = \kappa (\textbf{X}^*, \textbf{X}^*)$ and $\textbf{K}^{*} = \kappa (\textbf{X}, \textbf{X}^*)$. Thus, following the conditioning theorem for Gaussians, the posterior predictive distribution of $\textbf{f}^{*}$ is: 
\begin{equation}
	\textbf{f}^{*} |\textbf{X},\textbf{X}^{*},\textbf{f} \sim \mathcal{N}(\textbf{m},\textbf{C}), \quad\textbf{m} = \textbf{K}^{*\textnormal{T}}\textbf{K}_y^{-1}\textbf{y}, \quad\textbf{C} =\textbf{K}^{**} -\textbf{K}^{*\textnormal{T}}\textbf{K}_y^{-1}\textbf{K}^*.
\end{equation}

In this work, a GPR model $\hat{\vect{f}}$ is constructed from a set of training data where the predictors are the simulation parameters $\vect{\mu}$ and the outputs are the projection coefficients $\textbf{u}_{L}(\vect{\mu})$ computed from the HF solutions as:
\begin{equation}
	\boldsymbol\mu \rightarrow \hat{\textbf{u}}_L(\boldsymbol\mu) \approx \hat{\vect{f}}(\boldsymbol\mu) \quad \textnormal{trained from} \quad (\boldsymbol\mu_i, \textbf{V}^T \textbf{u}_h(\boldsymbol\mu_i)).
\end{equation}
Once trained, the model is used to predict the projection coefficients for any desired unseen value of the simulation parameters $\boldsymbol\mu^*$, i.e. $\hat{\textbf{u}}_L(\boldsymbol\mu^*)$. This allows recovering the FO solution as follows:
\begin{equation}
	\textbf{u}_{p}(\boldsymbol\mu^*) = \textbf{V} \hat{\textbf{u}}_L(\boldsymbol\mu^*) \approx  \textbf{u}_h(\boldsymbol\mu^*).
\end{equation}
The Matlab function $\code{fitrgp()}$ is used to construct the GPR models with the \textit{Matérn 5/2} kernel:
\begin{equation}
	\kappa(x_i, x_j) = \sigma_\kappa^2\exp\left(1 + \frac{\sqrt{5}r}{\sigma_l} + \frac{5r^2}{3\sigma_l^2} \exp\left(-\frac{\sqrt{5}r}{\sigma_l}\right)\right),
\end{equation}
where $r = \sqrt{(x_i-x_j)^\textnormal{T}(x_i-x_j)}$. Once trained, we can get the projection coefficients for any unseen combinations of predictors with the function \code{predictExactWithCov()}. The powerful aspect of GPR is that the output of the prediction is a multivariate normal distribution, characterised by a mean vector and a covariance matrix: $\mathcal{N}(\textbf{u}_{L}, \textbf{K}_{L})$. Therefore, the FO solution is also a multivariate normal distribution, $\mathcal{N}(\textbf{u}_{p} = \textbf{V}\textbf{u}_{L}, \textbf{K}_{p}=\textbf{V}\textbf{K}_{L}\textbf{V}^T)$. We can visualise the prediction uncertainty by sampling this distribution using the Matlab function \code{mvnrnd()}. Since in Matlab we are limited to IGPR, i.e. one GPR model for each projection coefficient, the covariance matrix is diagonal. 

\subsubsection{Metrics}
\noindent For the test cases, the ML-based ROM results are evaluated against the HF solutions by computing three absolute errors at each node of the stent mesh:
\begin{itemize}
	\item the order reduction error, $\text{E}_{rb} = ||\textbf{u}_{rb}(\boldsymbol\mu) - \textbf{u}_{h}(\boldsymbol\mu)||$;
	\item the prediction error, $\text{E}_{p} = || \textbf{u}_{p}(\boldsymbol\mu) - \textbf{u}_{h}(\boldsymbol\mu) ||$;
	\item the GPR error, $\text{E}_{gpr}  = \text{E}_{p} - \text{E}_{rb}$.
\end{itemize}

At the nodal level, these errors correspond to the distance between pairs of nodes. The accuracy of a single solution can be evaluated as the average error (AE) or maximum error (ME) on the mesh nodes: e.g. for the order reduction error, we get $\text{AE}_{rb}$ and $\text{ME}_{rb}$. Global variables can be computed by averaging these values on the test dataset: e.g. for the order reduction error, we refer to $\overline{\text{AE}}_{rb}$ as the average of $\text{AE}_{rb}$ over the test solutions and to $\overline{\text{ME}}_{rb}$ as the average of $\text{ME}_{rb}$ over the test solutions. 

\section{Results}\label{sec:results}

\begin{table}[ht!]
	\small
	\begin{center}
		\begin{tabular}{ccccccc}
			\hline
			$N_{s}$ & $N_{success}$ & $N_{failure}$ & $N_{train, class}$ & $N_{test, class}$ & $N_{train, regr}$ & $N_{train, regr}$ \\
			\hline
			150 & 97 & 53 & 50 & 100 & 47 & 50 \\
			300 & 196 & 104 & 200 & 100 & 146 & 50\\
			600 & 396 & 204 & 500 & 100 & 346 & 50\\
			900 & 605 & 295 & 800 & 100 & 555 & 50\\
			\hline
		\end{tabular}
		\caption{Number of train and test samples of the binary classifiers ($N_{train, class}$, $N_{test, class}$) and the ML-based ROM ($N_{train, regr}$, $N_{test, regr}$).}\label{tab:database_size} 
	\end{center}
\end{table}

\subsection{Binary classification: Prediction of deployment success}

In this section, the results of the classification models are presented. To study the influence of the dataset size on the prediction capability of the ML models, we first created a dataset of $N_s$ = 900 simulations and then, by using the \code{subLHCoptim()} function from the LatinHypercubeSampling.jl package, we defined three optimal subspaces with $N_s = \{$150,300,600$\}$. Following an approach similar to \citep{Hesthaven2018a}, we considered a fixed number of samples ($N_{test}$ = 100) for testing the four different datasets. In Table \ref{tab:database_size}, the number of samples considered for ML training and testing is reported. The input data were standardised before training.

\begin{figure}[ht!]
	\centering
	\includegraphics[width=0.9\textwidth]{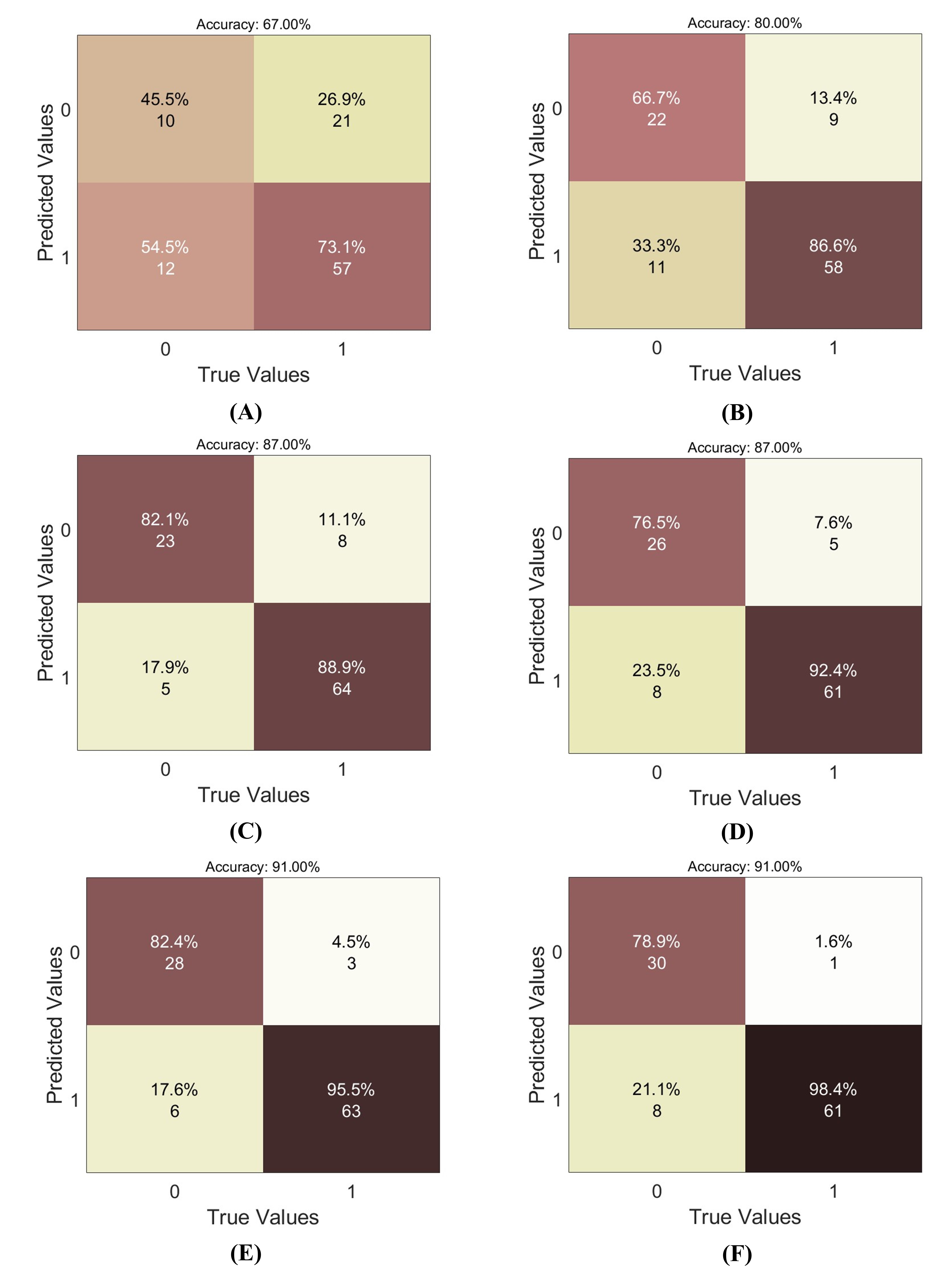}
	\caption{Analysis of classifiers  performance. Confusion matrices of the six ML models for $N_{train}$ = 200. \textbf{(A)} Logistic Regression. \textbf{(B)} k-Nearest Neighbour. \textbf{(C)} Naive Bayes. \textbf{(D)} Decision Tree. \textbf{(E)} Artificial Neural Network. \textbf{(F)} Support Vector Machines. }\label{fig:confmat}
\end{figure}

The values of the metrics for the four different dataset sizes are given in Table \ref{tab:class_results}. For $N_{train}$ = 200, the confusion matrices of the different classification models are reported in Figure \ref{fig:confmat}. The NB, ANN and SVM models all show an AUC larger than 96$\%$ when using $N_{train}$ = 200. The performance of all classification models improves when the training dataset is expanded. In general, the LR model shows the worst performance: regardless of the dataset size, its metrics stay below 90$\%$. Precision is especially low (highest value = 32.3$\%$): as visible in Figure \ref{fig:confmat}\textbf{A}, the model fails mostly in predicting false cases (failures). For the smallest dataset, the SVM model shows the best performance with accuracy, specificity and precision larger than 90$\%$. When considering $N_{train}$ = 500, all the validation metrics are larger than 90$\%$ for the SVM and ANN models. K-NN, DT and NB models also show good performance in terms of accuracy and specificity (between 83$\%$ and 96$\%$); however, when compared with SVM and ANN, they have poorer precision, which is reflected in a lower F1-score (maximum 83.3$\%$). The ANN and SVM models achieve the highest value of specificity (97$\%$), precision (93.9$\%$) and F1-score (92$\%$); instead, the highest sensitivity is reached by the NB model (95.8$\%$)). When considering the largest dataset $N_{train}$ = 800, only a few improvements are observed, e.g. the specificity of the NB model increases from 95.8$\%$ to 100$\%$. However, the majority of the metrics stabilise or even decrease in value: most likely, the models are undergoing overfitting. The results of the classification are related to the use of $\vect{\mu}_B$ as predictors; no improvement is observed using $\vect{\mu}_{cl}$ for classification.

\begin{table}[ht!]
	\small
	\begin{center}
		\begin{tabular}{ccccccc}
			\hline
			$N_s$ & Classification model & Accuracy & Sensitivity & Specificity & Precision & F1-score \\ \hline
			\multirow{6}{*} {150} & Logistic Regression & 75.0$\%$ & 87.5$\%$ & 73.9$\%$ & 22.5$\%$ & 35.8$\%$ \\
			&K-nearest neighbour & 86.0$\%$ & 81.4$\%$ & 87.6$\%$ & 70.9$\%$ & 75.8$\%$\\
			& Decision Tree & 89.0$\%$ & 100$\%$* & 86.2$\%$ & 64.5$\%$ & 78.4$\%$ \\
			& Naive Bayes & 89.0$\%$ & 45.4$\%$ & 70.5$\%$ & 30.3$\%$ & 36.3$\%$ \\
			& ANN & 80.0$\%$ & 61.7$\%$ & 96.2$\%$ & 93.5$\%$* & 74.3$\%$ \\
			& SVM*  & 93.0$\%$* & 85.2$\%$ & 96.9$\%$* & 93.5$\%$* & 89.2$\%$*\\
			\hline
			\multirow{6}{*} {300} & Logistic Regression & 67.0$\%$ & 45.4$\%$ & 73.0$\%$ & 32.2$\%$ & 37.7$\%$ \\
			& K-nearest neighbour & 80$\%$ & 66.6$\%$ & 86.5$\%$ & 70.9$\%$ & 68.7$\%$ \\
			& Decision Tree  & 87$\%$ & 82.1$\%$* & 88.8$\%$ & 74.1$\%$ & 77.9$\%$  \\
			& Naive Bayes & 87$\%$ & 76.4$\%$ & 92.4$\%$ & 83.8$\%$ & 80.0$\%$ \\
			& ANN*  & 91.0$\%$* & 78.9$\%$ & 98.3$\%$* & 96.7$\%$* & 86.9$\%$* \\
			& SVM & 90.0$\%$ & 80.0$\%$ & 95.3$\%$ & 90.3$\%$ & 84.8$\%$ \\
			\hline
			\multirow{6}{*} {600} & Logistic Regression  & 74$\%$ & 77.7$\%$ & 73.6$\%$ & 22.5$\%$ & 35.0$\%$ \\
			& K-nearest neighbour  & 83.0$\%$ & 73.3$\%$ & 87.1$\%$ & 70.9$\%$ & 72.1$\%$\\
			& Decision Tree  & 89.0$\%$ & 88.4$\%$ & 88.4$\%$ & 74.1$\%$ & 80.7$\%$ \\
			& Naive Bayes  & 91.0$\%$ & 95.8$\%$* & 89.4$\%$ & 74.1$\%$ & 83.6$\%$ \\
			& ANN*  & 95.0$\%$* & 90.6$\%$ & 97.0$\%$* & 93.5$\%$* & 92.0$\%$* \\
			& SVM*   & 95.0$\%$* & 90.6$\%$ & 97.0$\%$* & 93.5$\%$* & 92.0$\%$* \\
			\hline
			\multirow{6}{*} {900} & Logistic Regression & 75.0$\%$ & 87.5$\%$ & 73.9$\%$ & 12.1$\%$ & 19.5$\%$ \\
			& K-nearest neighbour & 86.0$\%$ & 81.4$\%$ &  87.6$\%$ & 70.9$\%$ & 75.8$\%$ \\
			& Decision Tree & 89.0$\%$ & 83.3$\%$ & 91.4$\%$ & 80.6$\%$ & 81.9$\%$ \\
			& Naive Bayes & 89.0$\%$ & 100.0$\%$* & 86.2$\%$ & 64.5$\%$ & 78.4$\%$ \\
			& ANN* & 94.0$\%$* & 87.8$\%$ &97.0$\%$* &  93.5$\%$* & 90.6$\%$* \\
			& SVM   & 93$\%$ & 85.2$\%$ & 96.9$\%$ & 93.5$\%$* & 89.2$\%$   \\
			\hline
		\end{tabular}
		\caption{\label{tab:class_results} Analysis of classifiers performance. Evaluation metrics of the six ML models for $N_{train} = \{$50,200,500,800$\}$. The best result for each category is highlighted. }
	\end{center}
\end{table}

\subsection{Reduced order modelling: Prediction of the stent deployed configuration}

In this section, the results of the ML-based ROM are presented. As proposed for the classification models, we started with a dataset with $N_s$ = 900 simulations and then defined three optimal subspaces with $N_s = \{$150, 300, 600$\}$. Then, we built four datasets with only the successful deployment simulations, $N_{success} = \{$97, 196, 396, 555$\}$, and considered a fixed number of samples ($N_{test}$ = 50) for testing. In Table \ref{tab:database_size}, the number of samples considered for ML-based ROM training and testing is reported. The input and output data were standardised before training. The spatial resolution of imaging techniques currently used for IAs detection and treatment was considered for evaluating the prediction errors \citep{Hacein-Bey2011, Maupu2022}: magnetic resonance angiography (MRA) = 0.6–1mm, computed tomography angiography (CTA) = 0.4–0.7 mm, digital subtraction angiography (DSA) = 0.2 mm, 3D rotational angiography (3DRA) = 0.15 mm.

\begin{figure}[ht!]
	\centering
	\includegraphics[width=\textwidth]{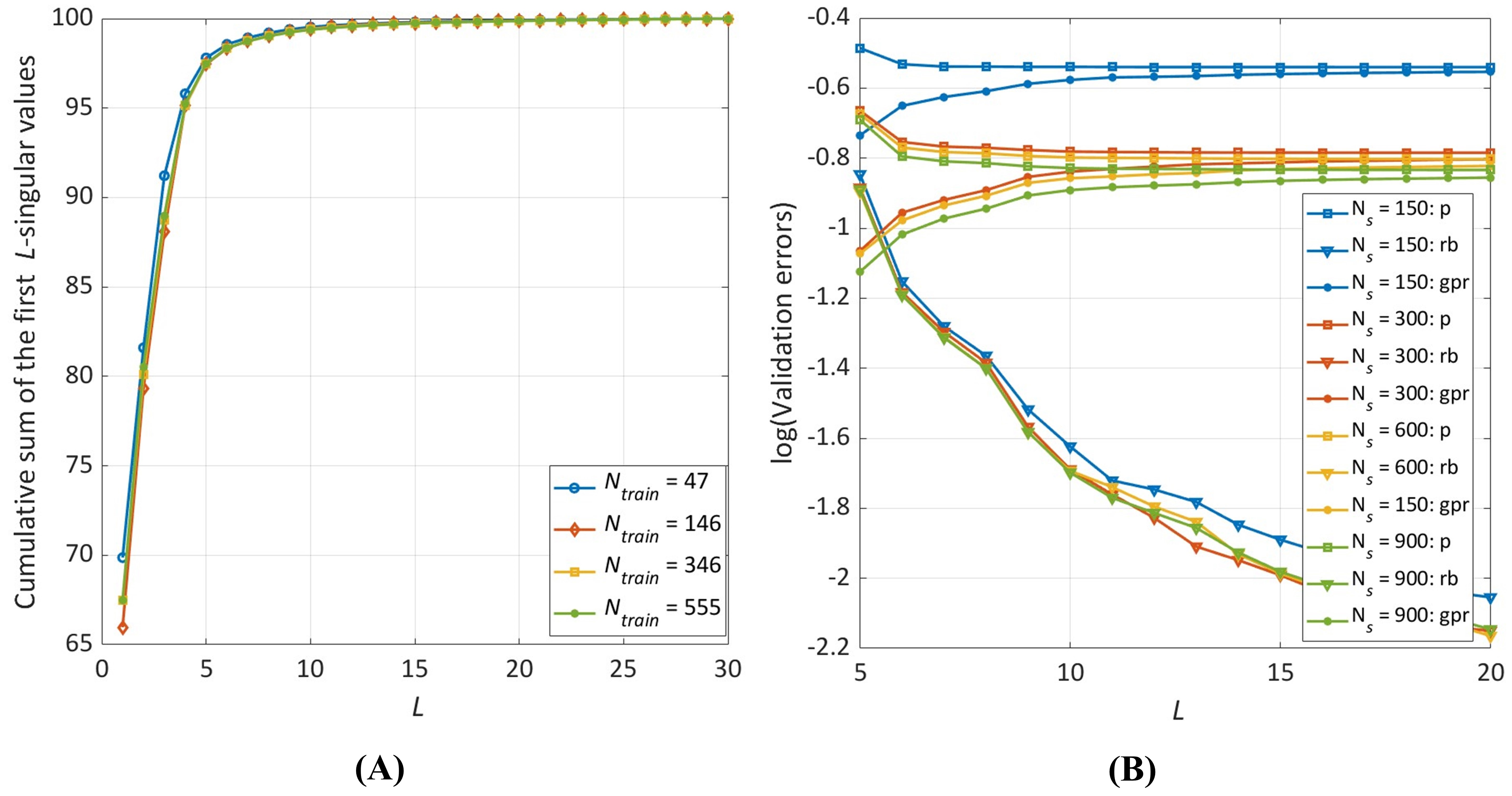}
	\caption{Sensitivity analysis on the number of RBs ($L$) and the training database size ($N_{train}$). The results are obtained using the Bézier curve parameters ($\vect{\mu}_{B}$) as GPR predictors. \textbf{(A)} Percentage of variance of the snapshots matrix ($\vect{S}$) explained by the first $L$ components of the SVD. \textbf{(B)} Evolution in logarithmic scale of $\overline{\text{AE}}_{rb}$, $\overline{\text{AE}}_{p}$ and $\overline{\text{AE}}_{gpr}$. } \label{fig:err_L}
\end{figure}

As already mentioned in Section \ref{sec:rom}, the choice of the RBs number $L$ is carried out considering the singular values of the snapshots matrix. In Figure \ref{fig:err_L}\textbf{A}, the cumulative sum of the singular values normalised with respect to their total sum is reported: this plot shows the fraction of total variance retained by the first $L$-singular values for $N_{train} = \{$47,146,346,555$\}$. As reported in Table \ref{tab:errs_ns}, regardless of the dataset size, the first 4 RBs cover more than 90$\%$ ($\varepsilon_{POD}$ = 0.1) and the first 10 RBs more than 99$\%$ ($\varepsilon_{POD}$ = 0.01) of the total variance in the input data. As shown in Figure \ref{fig:err_L}\textbf{B}, the prediction does not benefit considerably by considering a number of RBs greater than 15: in fact, analogously for any value of $N_{train}$, the average order reduction error decreases towards 0 as more RBs are considered, while the average prediction and GPR errors reach a stable plateau around $L$ = 15. Therefore, henceforth $L$ = 15 is considered in this analysis. 

\begin{figure}[ht!]
	\centering
	\includegraphics[width=\textwidth]{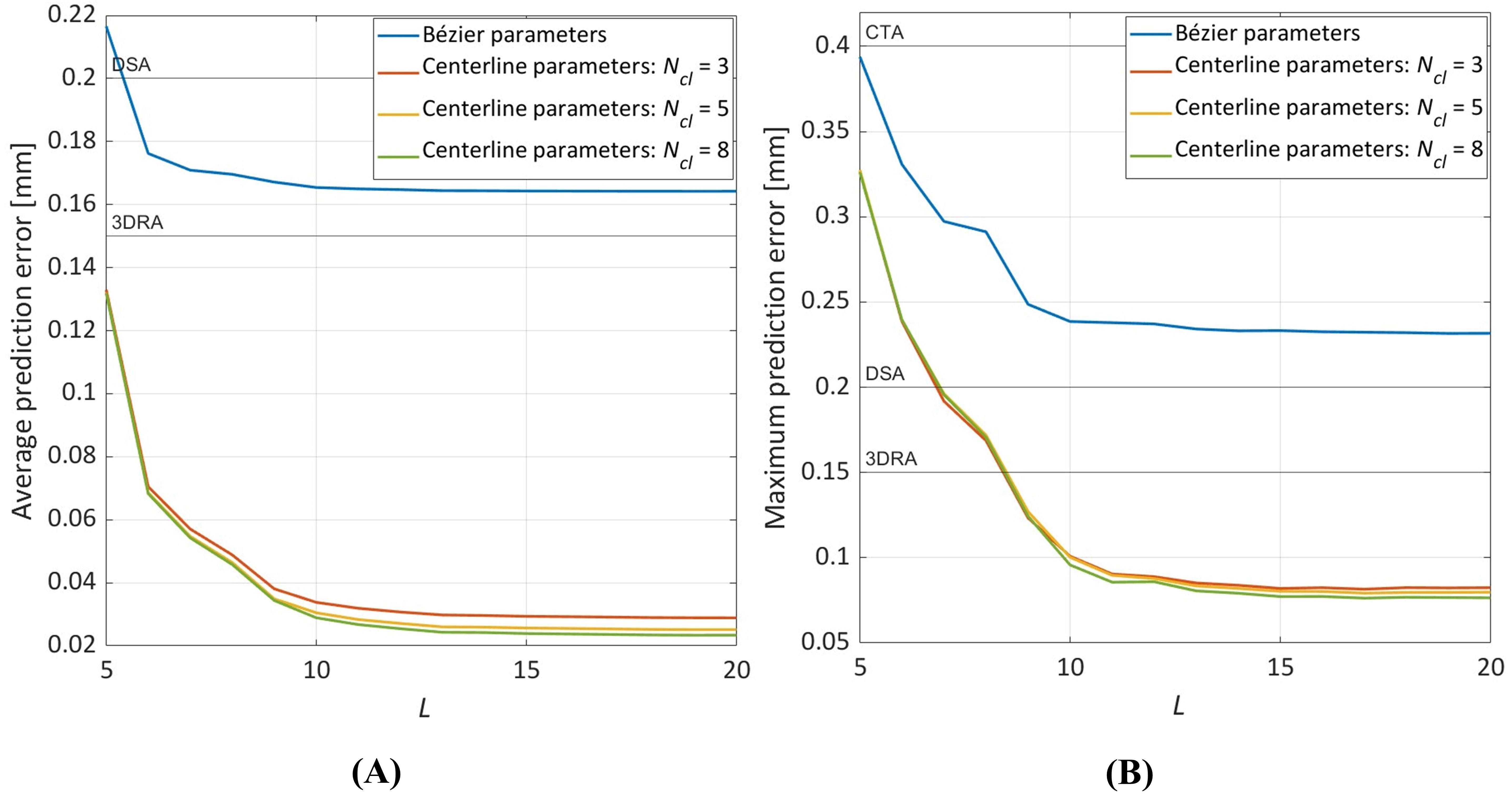}
	\caption{Sensitivity analysis on the parameters used as GPR predictors: Bézier curve parameters ($\vect{\mu}_{B}$) vs centerline points parameters 	($\vect{\mu}_{cl}$). The results are obtained with $L$ = 15 and $N_{train}$ = 146. \textbf{(A)} Evolution of $\overline{\text{AE}}_{p}$. \textbf{(B)} Evolution of $\overline{\text{ME}}_{p}$. } \label{fig:err_cl}
\end{figure}

\begin{figure}[ht!]
	\centering
	\includegraphics[width=\textwidth]{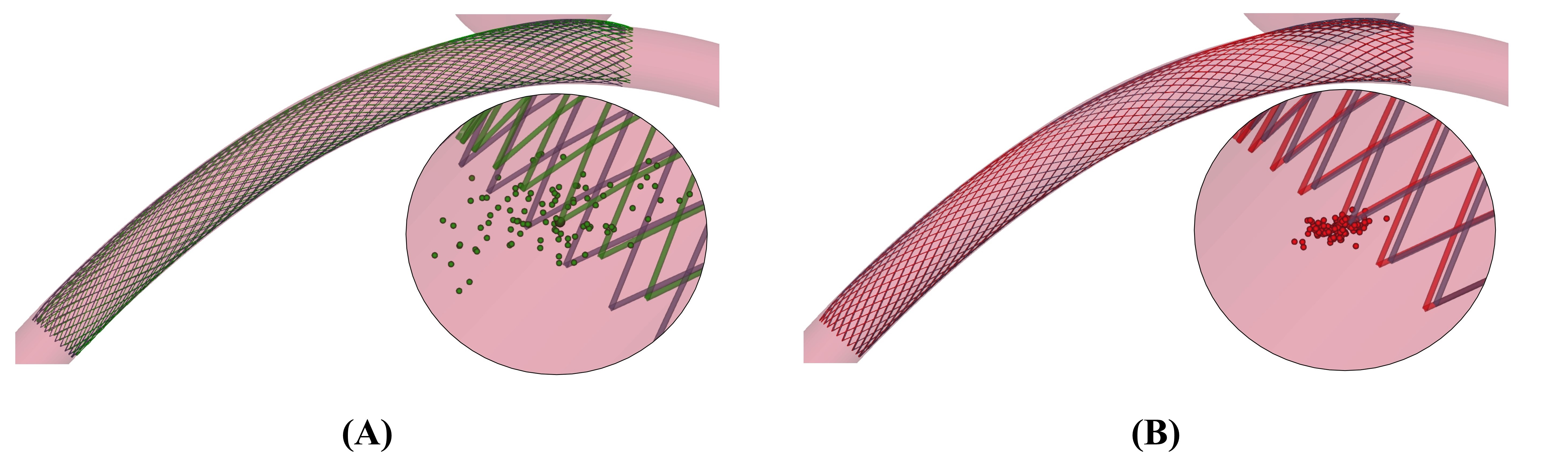}
	\caption{Two examples from the test dataset: focus on 100 points sampled within the multivariate normal distribution predicted by the GPR for the displacement of the first node of the stent mesh. \textbf{(A)} HF (black) and predicted (green) solution using $\vect{\mu}_B$ as predictors. \textbf{(B)} HF (black) and predicted (red) solution using $\vect{\mu}_{cl}$ as predictors.}\label{fig:scatter}
\end{figure}

The results presented so far relate to the use of $\vect{\mu}_B$ as predictors. As reported in Table \ref{tab:errs_cl} and shown in Figure \ref{fig:err_cl}, a strong reduction of the prediction error is evident when using $\vect{\mu}_{cl}$ instead: with $N_{cl}$ = 3, the average prediction error is 5.65$\times$ lower than when using $\vect{\mu}_B$ while the maximum prediction error is 2.84$\times$ lower. Increasing the number of considered points $N_{cl}$, a further but slight decrease of the prediction error is observed: the average and maximum values are respectively 1.11$\times$ and 1.03$\times$ lower with $N_{cl}$ = 5 and 1.08$\times$ and 1.04$\times$ lower with $N_{cl}$ = 8. Using $\vect{\mu}_{cl}$, the percentage of test samples where the average prediction error is greater than the spatial resolution of 3DRA is zero; the same is true for the number of test samples where the maximum prediction error is greater than the spatial resolution of CTA. The number of test samples where the maximum prediction error is greater than the spatial resolution of 3DRA is reduced from 64$\%$ with $\vect{\mu}_B$ to 10$\%$ with $\vect{\mu}_{cl}$. The improvement gained by using $\vect{\mu}_{cl}$ is not only reflected in a lower variance of the prediction error within the test dataset but also within a single test solution: this is noticeable when sampling the multivariate normal distributions corresponding to the GPR outputs, as explained in \ref{sec:gpr}. In Figure \ref{fig:scatter}, we reported the position of 100 points sampled within the distribution predicted by the GPR for the displacement of the first node of the stent mesh. The points are more widely distributed with $\vect{\mu}_B$ while they are highly concentrated around the mean node value with $\vect{\mu}_{cl}$, which means that the uncertainty on the predicted displacement value when using $\vect{\mu}_{cl}$ is lower than when using $\vect{\mu}_B$. 

\begin{figure}[h!]
	\centering
	\includegraphics[width=\textwidth]{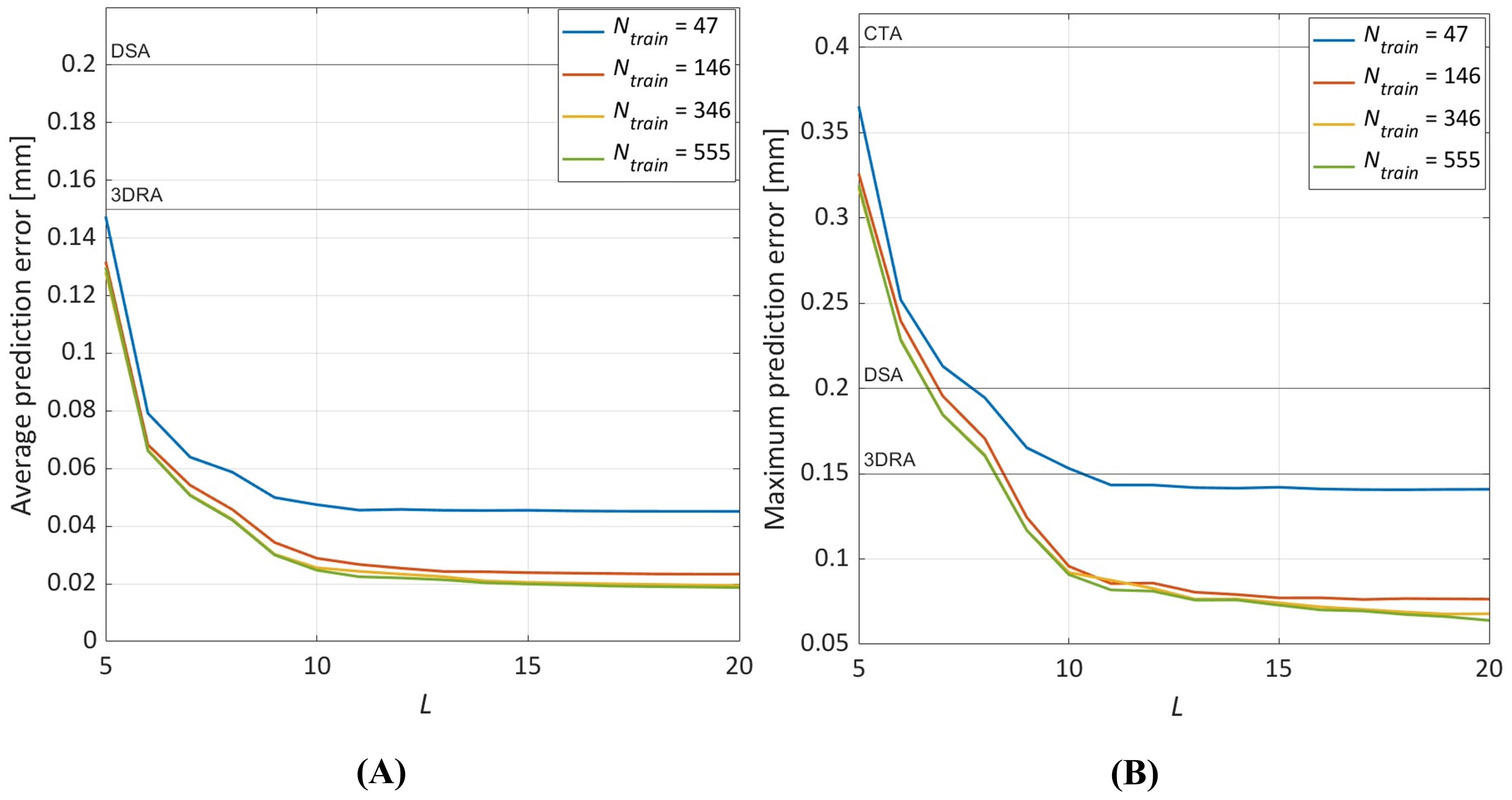}
	\caption{Sensitivity analysis on the training database size ($N_{train}$). The results are obtained with $L$ = 15 and using the centerline points parameters ($\vect{\mu}_{cl}$ with $N_{cl}$ = 8) as GPR predictors. \textbf{(A)} Evolution of $\overline{\text{AE}}_{p}$. \textbf{(B)} Evolution of $\overline{\text{ME}}_{p}$. } \label{fig:err_ns}
\end{figure}

Finally, we analysed the influence of the number of train samples on the ML-based ROM performance. Considering $N_{train}$ = 47 instead of $N_{train}$ = 146, the average prediction error is 1.92$\times$ lower while the maximum prediction error is 1.84$\times$ lower. The average and maximum prediction error slightly decrease when considering $N_{train}$ = 346 instead of $N_{train}$ = 146, respectively 1.14$\times$ and 1.04$\times$. It can be observed in Figure \ref{fig:err_ns} that further expanding the training dataset, the prediction error converges around the value reached when $N_{train}$ = 346 is used. 

\begin{table}[ht!]
	\small
	\begin{center}
		\begin{tabular}{ccccccc}
			\hline
			$N_{train}$ & $\varepsilon_{POD}$ & $L$  &  $\overline{\text{AE}}_{rb}\pm$SD  &
			$\overline{\text{ME}}_{rb}\pm$SD  & $\overline{\text{AE}}_{p}\pm$SD   & $\overline{\text{ME}}_{p}\pm$SD\\
			& & & [mm] & [mm] & [mm] & [mm]  \\
			\hline
			\multirow{3}{*} {47} & 0.1 & 3 & 0.730$\pm$0.588 & 1.797$\pm$1.797  & 0.813$\pm$0.628  & 1.842$\pm$1.343 \\& 0.01 & 8 & 0.043$\pm$0.025 & 0.171$\pm$0.097  & 0.289$\pm$0.238  & 0.460$\pm$0.324 \\ & 0.001 & 25 & 0.008$\pm$0.004 & 0.036$\pm$0.025  & 0.288$\pm$0.239  & 0.437$\pm$0.325 \\
			\hline
			\multirow{3}{*} {146} & 0.1 & 4 & 0.271$\pm$0.168 & 0.728$\pm$0.483  & 0.331$\pm$0.189  & 0.757$\pm$0.429 \\& 0.01 & 9 & 0.027$\pm$0.016 & 0.111$\pm$0.065  & 0.167$\pm$0.119  & 0.249$\pm$0.146 \\ & 0.001 & 58 & 0.003$\pm$0.002 & 0.014$\pm$0.008  & 0.164$\pm$0.120  & 0.232$\pm$0.147 \\
			\hline
			\multirow{3}{*} {346} & 0.1 & 4 & 0.271$\pm$0.166 & 0.723$\pm$0.469 & 0.329$\pm$0.182  & 0.746$\pm$0.413 \\& 0.01 & 10 & 0.020$\pm$0.010 & 0.078$\pm$0.042 & 0.159$\pm$0.119  & 0.217$\pm$0.130 \\& 0.001  & 99 & 0.002$\pm$0.001 & 0.008$\pm$0.004  & 0.157$\pm$0.120  & 0.198$\pm$0.131 \\
			\hline
			\multirow{3}{*} {555} & 0.1 & 4 & 0.270$\pm$0.161 & 0.724$\pm$0.463 & 0.323$\pm$0.172 & 0.744$\pm$0.403 \\ & 0.01 & 10 & 0.020$\pm$0.010 & 0.077$\pm$0.041 & 0.148$\pm$0.111 & 0.204$\pm$0.121 \\&  0.001 & 124 &0.001$\pm$0.001 & 0.006$\pm$0.003  & 0.146$\pm$0.112 & 0.181$\pm$0.121 \\
			\hline
		\end{tabular}
		\caption{\label{tab:errs_ns} Analysis of ML-based ROM performance considering different tolerances ($\varepsilon_{POD}$), numbers of RBs ($L$) and training database sizes ($N_{train}$). The results are obtained using the Bézier curve parameters ($\vect{\mu}_B$) as GPR predictors.}
	\end{center}
\end{table}
\begin{table}[ht!]
	\small
	\begin{center}
		\begin{tabular}{lccccc}
			\hline
			Predictors & $\overline{\text{AE}}_{p}\pm$SD & $N_{\text{AE}_p> 0.15\text{mm} = \text{3DRA}}$  & $\overline{\text{ME}}_{p}\pm$SD & $N_{\text{ME}_p> 0.15\text{mm} = \text{3DRA}}$ & $N_{\text{ME}_p> 0.4\text{mm} = \text{CA}} $  \\
			& [mm] &  & [mm]  &  &  \\
			\hline
			Bézier & 0.164$\pm$0.120 & 26(52$\%$) & 0.233$\pm$0.147 & 32(64$\%$) & 4(8$\%$) \\
			$N_{cl} = 3$ & 0.029$\pm$0.020 & 0(0$\%$) & 0.082$\pm$0.058 & 5(10$\%$) & 0(0$\%$) \\
			$N_{cl} = 5$ & 0.026$\pm$0.017 & 0(0$\%$) & 0.080$\pm$0.059 & 5(10$\%$) & 0(0$\%$) \\
			$N_{cl} = 8$ & 0.024$\pm$0.017 & 0(0$\%$) & 0.077$\pm$0.058 & 6(12$\%$) & 0(0$\%$)\\
			\hline
		\end{tabular}
		\caption{\label{tab:errs_cl} Analysis of ML-based ROM performance comparing the Bézier curve parameters ($\vect{\mu}_B$) and the centerline points parameters ($\vect{\mu}_{cl}$) as GPR predictors. The results are obtained with $L$ = 15 and $N_{train}$ = 146.}
	\end{center}
\end{table}

\section{Discussion}\label{sec:disc}
In this work, a fast and accurate method for braided stent deployment analysis has been proposed. It consists of a two-step workflow where FE simulations are used, first, to train a ML model that classifies successful and unsuccessful simulations and, next, to train a ML-based ROM that approximates the stent deployed configuration within the considered vessel. This approach was validated by studying the effect of a combination of geometrical and surgical parameters on the outcome of the stent deployment simulation: to this end, we employed a parametric idealised model of an intracranial artery characterised by a saccular aneurysm. The presented model relies on previous work to develop an optimised framework for numerical simulations of frictional contact interactions between wire-like structures discretised using beam elements and rigid surfaces. Based on this, here we proposed an efficient approach to simulate braided stent deployment, which allowed us to reduce the computational time required to perform the simulations needed for ML training. With the scheme proposed in Section \ref{sec:hf}, the computational time to perform a FE simulation of stent deployment takes, on average, 15 minutes. This makes it possible to build even large datasets in an acceptable amount of time: e.g., the largest dataset $N_s$ = 900 used in this analysis could be created in 12h on 20 nodes of a cluster with 2.6 Ghz Intel Xeon Gold 6132 CPUs and 6 Gb of RAM for each node.

\begin{figure}[ht!]
	\centering
	\includegraphics[width=0.9\textwidth]{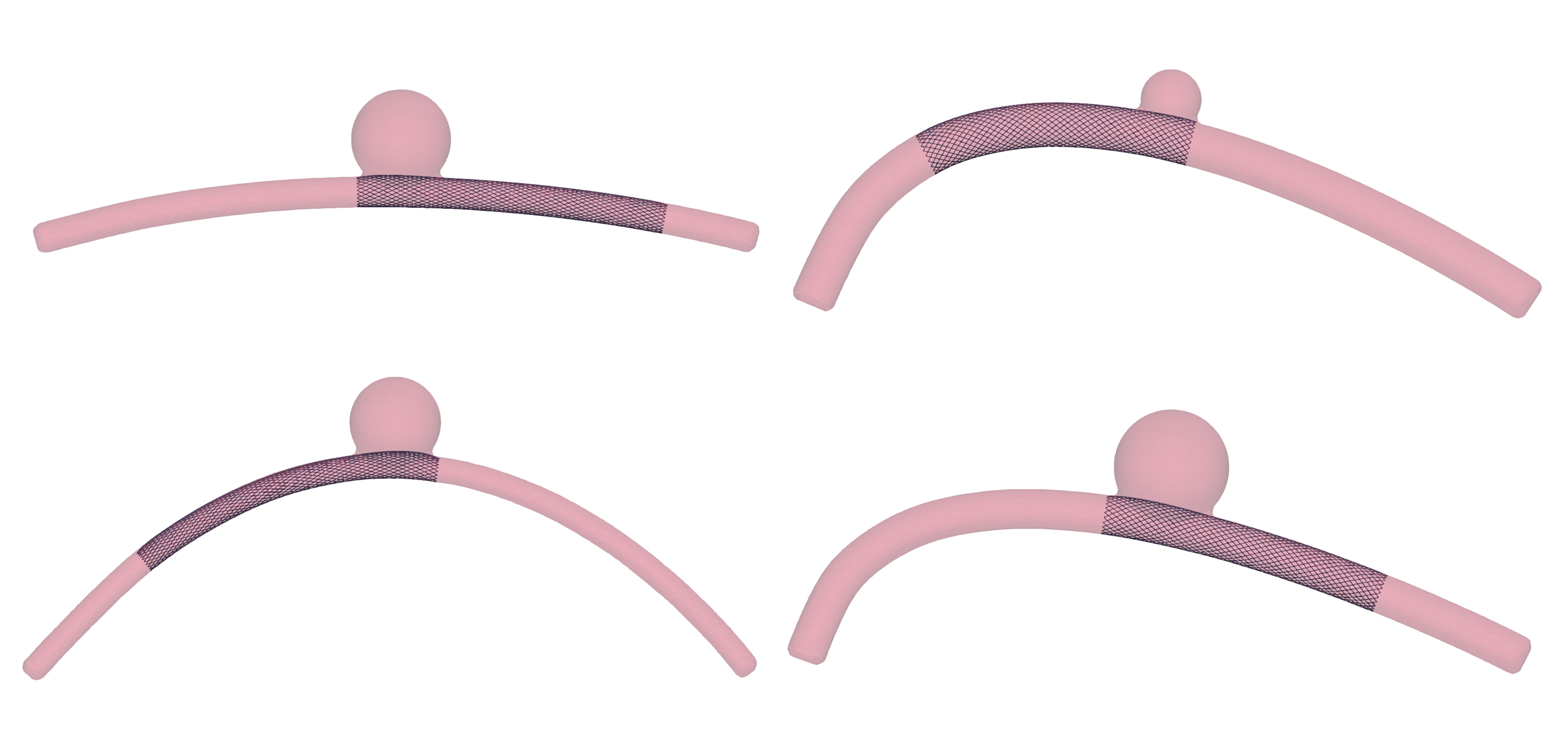}
	\caption{\label{fig:fn} Four examples of false negatives misclassification from the test dataset.} 
\end{figure}

We trained six ML models on a dataset of varying size and obtained classifiers that were 80-91$\%$ accurate in predicting the deployment outcome even with a relatively small dataset ($N_{train}$ = 146). Increasing its size to $N_{train} = 346$ and using SVM and NN models, we were able to perform binary classification with all the validation metrics between 89-97$\%$. The surgical needs addressed by this model require us to favour the presence of FNs over that of FPs, i.e., we prefer to mislabel a successful simulation as a failure rather than the reverse. To minimize the presence of FPs then, high-specificity models are preferred. Therefore, for $N_{train}$ = 47, we would select the SVM model and for $N_{train}$ = 146, the ANN model. As shown in Figure \ref{fig:fn}, FNs misclassification may be explained by the fact that, in these cases, the stent is in a "boundary" situation in which one of its extremities lands immediately after the aneurysm neck. Introducing deformable walls, these cases would most probably fall into the "failure" condition. A similar situation is observed also for some FPs cases, i.e. one of the extremities of the stent lands within the aneurysm immediately before the aneurysm neck, but no such clear explanation was found for all the FPs.

\begin{figure}[ht!]
	\centering
	\includegraphics[width=\textwidth]{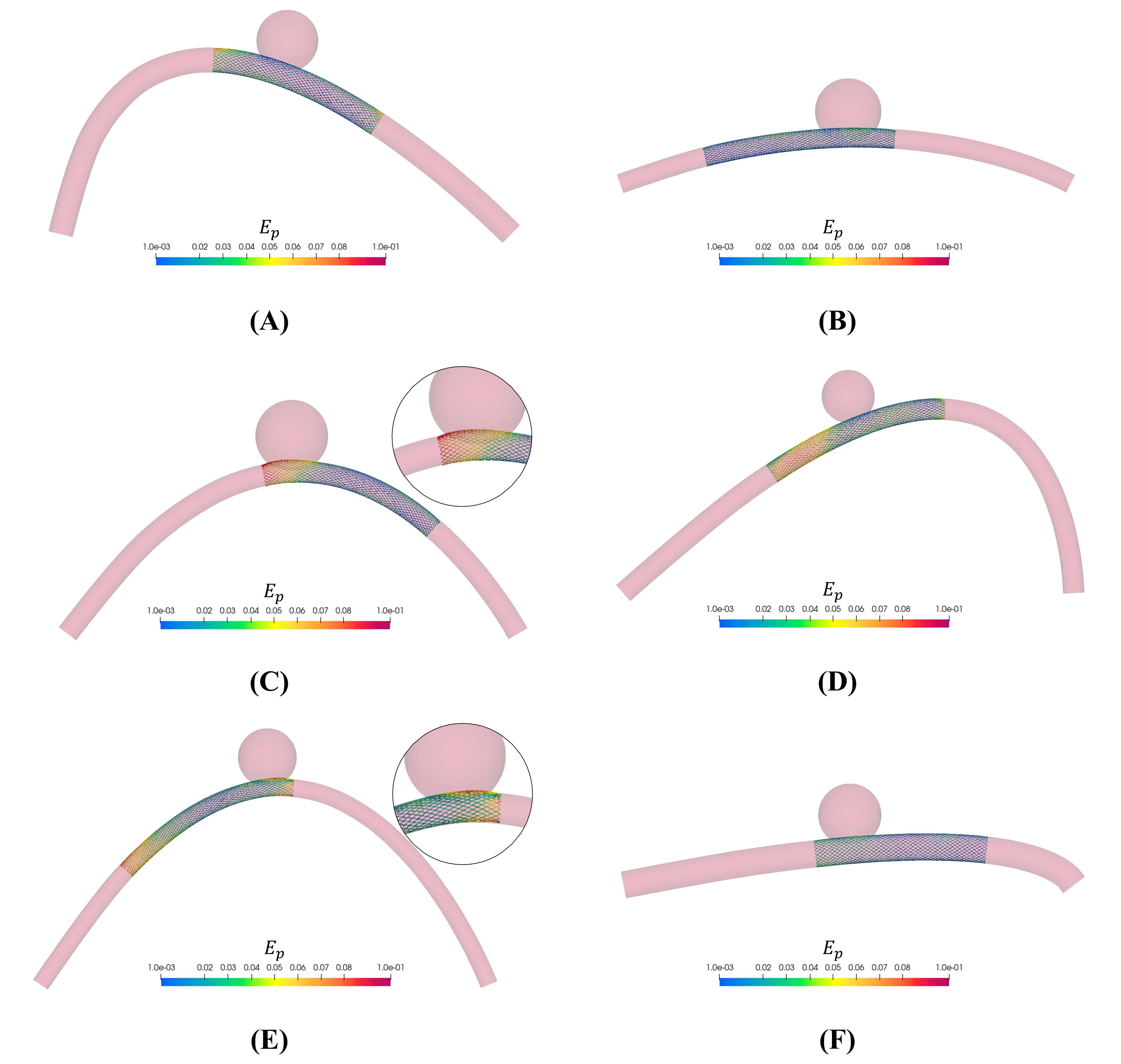}
	\caption{Six examples from the test dataset. The nodal absolute error $E_p$ between HF and predicted solution is shown as a colourmap. As comparative scale, for each solution we reported the diameter of the vessel and the aneurysm. The wire thickness is magnified (2$\times$) to better visualise the stent. \textbf{(A)} $D_{v}$ = 3.3 mm, $D_{a}$ = 8.24 mm. \textbf{(B)} $D_{v}$ = 2.62 mm, $D_{a}$ = 8.84 mm. \textbf{(C)} $D_{v}$ = 2.9 mm, $D_{a}$ = 9.68 mm. \textbf{(D)} $D_{v}$ = 2.84 mm, $D_{a}$ = 7.08 mm. \textbf{(E)} $D_{v}$ = 2.38 mm, $D_{a}$ = 7.86 mm.  \textbf{(F)} $D_{v}$ = 3.62 mm, $D_{a}$ = 8.4 mm.} \label{fig:regr_examples}
\end{figure}

For the regression, the POD algorithm was employed to extract the RBs while a GPR model to establish a mapping between the simulation parameter values and the projection coefficients. The results showed that GPR is strongly influenced by the input parameters: maintaining the same output data and changing the predictors from $\vect{\mu}_{B}$ to $\vect{\mu}_{cl}$, the average prediction error could be decreased by more than 5$\times$. This reduction can be explained by the less non-linear relationship present between these alternative predictors and the output variables. We carried out a sensitivity analysis on the number of RBs to be considered for the dimensionality reduction of the problem and on the training dataset size. The results showed that the prediction error stabilises with a relatively low number of RBs (15). On the other hand, with 47 training samples, we were able to achieve a maximum prediction error slightly lower than the spatial resolution of 3DRA; with 147 training samples, we were able to reduce it to 0.07 mm, half the spatial resolution of 3DRA. In our analysis, the average prediction error is never greater than 0.15 mm (3DRA spatial resolution) while the maximum prediction error is never greater than 0.4 mm (CTA spatial resolution) and is lower than 0.15 mm (3DRA spatial resolution) in 90$\%$ of the total test cases. By looking at the test examples reported in Figure \ref{fig:regr_examples}, it can be observed that the prediction error is evenly distributed and the stent configuration accommodates very well the vessel curvature. In light of these results, we can conclude that a prediction error in the range of 0.01-0.05 mm leads to an approximated stent that is very close to the HF results and such an error is achievable even with a relatively small dataset (47 simulations). In the cases characterised by larger prediction errors (Figure \ref{fig:regr_examples}\textbf{C} and \ref{fig:regr_examples}\textbf{E}), the greatest differences can be observed at the stent extremities. These cases correspond to the same "boundary" situation that also affects the classification models, where one of the stent extremities lands immediately after the aneurysm neck. As mentioned above, by introducing deformable walls, the stent would most likely slip within the aneurysmal space in this situation, making the deployment unsuccessful. Thanks to the complete decoupling between the offline stage (extraction of RBs and training of the regression model) and the online one (prediction of new solutions), the computational cost is decreased from $\sim$15 minutes, using the FE simulation scheme proposed, to a few milliseconds. 

The results here presented are promising as they demonstrate the ability of ML and reduced order modelling techniques to account for the non-linearities of the stent deployment problem and accurately model its outcome. However, this approach presents some limitations that need to be addressed. First, the main limitation is the ability to consider only one stent geometry at a time: in fact, changes to the stent mesh corresponding to different amounts of DOFs would result in incompatible displacement vectors to be assembled in the snapshot matrix $\textbf{S}$. Therefore, considering stents of various sizes would require the construction of separate ROMs. Nevertheless, flow diverters are available in a finite and limited number of sizes, so creating different models could be feasible. Besides, the IGPR method used in this project does not model the correlation between outputs: alternatively, a multi-output GPR approach should be considered in the future \citep{Liu2018}. Secondly, both the stent and vessel FE models here considered are based on some simplifications: in particular, the rigid-wall hypothesis for the vessels and the penalty-based constraints at the wires interconnections for the stent. Deformable models and a more general beam-to-beam contact formulation would increase the computational cost of the HF simulations, introduce a higher level of complexity in the modelled problem and, thus, require a larger training dataset. Moreover, deformable walls would most probably change the outcome of the deployment simulation, in particular in the "boundary" situations discussed above. Finally, the vessel geometry considered for this work is rather far from patient-specific geometries. Introducing more parameters will result in the need for a larger number of bases (hence, a larger training dataset) to predict the solution well. Therefore, efficient algorithms to construct the training dataset should be explored to reduce its size as much as possible as well as alternative ML algorithms for regression, e.g. neural networks. 

\section{Conclusions}\label{sec:conclusions}
This work represents the first attempt to combine finite element simulations with machine learning and reduced order modelling for the analysis of braided stent deployment. Its feasibility was demonstrated using an idealised vessel model, where a set of geometrical features can be controlled. Surgical decisions were also taken into account in the creation of the high-fidelity dataset. The two-step workflow allows the classification of deployment conditions with up to 95$\%$ accuracy and real-time prediction of the stent deployed configuration with a maximum prediction error always lower than the spatial resolution of computed tomography angiography (0.4 mm) and lower than that of 3D rotational angiography (0.15 mm) in 90$\%$ of test cases. Despite the simplified vessel shape and the assumption of rigid walls, this study is representative of the clinical scenario and can be extended to more realistic applications without modification. Current efforts are focused on understanding how many parameters are needed to fully describe patient-specific models. To represent such geometries in the reduced-order model, a statistical shape model will be used instead of the parametrization presented here. In the future, a similar computational tool could be used by practitioners before and during intracranial aneurysm surgery to rule out conditions that would lead to unsuccessful deployment, visualize the stent configuration depending on the deployment site and check whether the chosen device covers one or more side branches.

\section{Funding and Acknowledgments}
This project is carried on in the framework of the MeDiTaTe Project, which has received funding from the European Union’s Horizon 2020 research and innovation programme under Grant Agreement 859836. Miquel Aguirre work has been partially supported by a Maria Zambrano research fellowship at Universitat Politècnica de Catalunya funded by Ministerio de Universidades. 

\newpage
\bibliographystyle{elsarticle-num}
\bibliography{manuscriptBB23}

\end{document}